# An Interactive Graphical Tool to Check the Coarray Continuity of Two-Fold Redundant Sparse Arrays (TFRSAs) Under Single Sensor Failures

Namya Malik, Ashish Patwari*, Sangeetha N

School of Electronics Engineering, Vellore Institute of Technology, Vellore, Tamil Nadu, India 632014

*Corresponding Author: ashish.p@vit.ac.in

https://orcid.org/0000-0001-9489-7004

*Abstract*—Two-fold redundant sparse arrays possess inbuilt redundancy to tackle single-element failures. This property enables them to perform accurate direction of arrival (DOA) estimation even during single sensor faults. However, recent literature suggests that some TFRSAs suffer from hidden dependencies whereby a single sensor fault at peculiar positions within the array cause discontinuities (holes) in the difference coarray (DCA). This violates the very idea of providing two-fold redundancy. Such hidden dependencies could prove catastrophic in many critical applications such as defense, autonomous driving, and biomedical imaging. Despite this issue, no formal tools or techniques exist to ascertain whether a given array configuration is truly twofold redundant or not. To address this gap, we provide a comprehensive framework and a first-ever graphical user interface (GUI). The GUI has been built using the features in MATLAB app designer and tested with known examples available in sparse array literature. Several numerical examples have been discussed to check the tool's response in each scenario. We conclude that the GUI is functionally accurate and can be an indispensable tool for sparse array designers in making informed choices about array configurations prior to real deployment.

## 1. Introduction

A sensor array contains at least two sensors arranged in a particular layout such as linear, circular, planar (V-shaped, L-Shaped, square layout, rectangular layout), spherical and so on [1], [2]. Sensor arrays are fundamental to spatial signal processing systems such as radar, sonar, wireless communications, and acoustic sensing as they possess directional properties unlike individual sensors. A major application of sensor arrays is to estimate the directions of incoming wave fields [3], [4], [5]. Sparse arrays gained tremendous traction in the past decade due to their ability to provide apertures (angular resolutions) comparable to that of regular/filled arrays using far fewer sensors [6]. Compared to full arrays, sparse arrays offer several advantages, including reduced hardware complexity, lower mutual coupling, and overall cost efficiency; making them attractive for practical systems [7].

Sparse arrays are governed by the difference coarray (DCA), which represents the set of all possible sensor separations in the physical array. When the DCA is continuous or hole-free, signal processing can be performed in the coarray domain. This gives the array an ability to resolve many more sources than available sensors [8]. Coarray processing has driven extensive research into efficient sparse array design [9], [10], [11], [12]. An array with hole-free DCA is also known as a *restricted array* in interferometric terminology [13], [14]. The DCA itself is known as *baseline*. Modern sparse array designs include nested arrays, coprime arrays, minimum redundancy arrays (MRAs), and various variations thereof [15], [16], [17], [18], [19], [20], [21], [22], [23], [24]. The array with maximum interelement spacing constraint (MISC) can be considered as a benchmark for all modern sparse arrays.

*1.1 Sensor Failures in Sparse Arrays and the Need for Two-Fold Redundancy*

Sensor failures in sparse arrays are a major concern. An array in which all sensors are non-redundant is said to be maximally economic. In other words, the correct functioning of each sensor is essential to preserve the hole-free property of the DCA. Unfortunately, most modern sparse arrays are maximally economic. Although optimal in terms of aperture efficiency, they lack the robustness to withstand even a single sensor failure, without affecting their DOA estimation ability [25].

On the other hand, multifold redundant sparse arrays rely on multi-level coarray redundancy to tackle sensor failures. In particular, two-fold redundant sparse arrays (TFRSAs) have inbuilt coarray redundancy to handle single sensor failures. In such arrays, each non-zero lag is generated by at least two distinct sensor pairs. Hence, if one sensor fails, the lag can be obtained from the other sensor pair. This ensures that the failure of any single sensor does not create holes in the coarray, thereby preserving functional performance. Several TFRSA formulations such as the symmetric nested array, the composite singer array, the 2FRA, the ULA-inspired TFRA have been proposed in the existing literature [26], [27], [28], [29]. However, not all TFRSAs live up to their promise of being robust to single sensor failures owing to hidden vulnerabilities that disrupt their hole-free coarray property under some specific scenarios, as described next.

*1.2 Motivation for a New GUI Tool To Check TFRA Robustness*

Although certain TFRAs appear structurally redundant, they might be prone to hidden dependencies that can disrupt coarray continuity during single sensor failures [30]. Recent studies have shown that some TFRA configurations consist of hidden essential sensors (HESs), whose failure can create holes in the DCA [29], [31]. This happens due to shared sensor redundancy i.e., due to the involvement of a single sensor in more than one instance of generating a given spatial lag. For instance, if an array has two sensor pairs, namely (2, 7) and (7, 12), that can generate a spatial lag 5, the sensor at {7} is said to be a HES as its failure leaves no option to obtain lag 5. This creates a hole in the DCA. Although lag 5 has a nominal multiplicity of two, both its occurrences are linked to the health of the sensor at {7}. Hence, the failure of the specific sensor at {7} disrupts the DCA continuity, making it an essential sensor. Detecting such fragility is non-trivial and requires systematic failure analysis. As shown later, the presence (or rather the failure) of a HES can severely degrade the array's DOA estimation capabilities, leading to ghost peaks or missed targets which are undesirable in applications such as radar, acoustics, biomedical imaging and so on.

Losing the hole-free coarray property during a single sensor failure dilutes the whole idea of providing two-fold redundancy in the first place. Hence, care must be exercised to ensure that the coarray is not only structurally redundant but also functionally robust. Manual verification could be error-prone, and purely theoretical checks may overlook hidden structural dependencies. If arrays with hidden vulnerabilities are unknowingly employed in real systems, they could cause undesired consequences. *Hence, there is a need for a systematic and practical procedure to check hidden dependencies in TFRAs. The tool shall immediately reveal whether the test array provides true robustness or just nominal redundancy.*

*1.3 Uniqueness of the proposed GUI*

While there are many array tools focusing on radiation pattern analysis and sidelobe analysis [32], [33], [34], there are practically no tools for sparse array analysis barring the recently introduced coarray analyzer [35]. The coarray analyzer can compute DCAs, plot weight functions, and verify the hole-free status of the coarray but it cannot perform the in-depth failure analysis required to ascertain array robustness. *To address this gap, we propose a*

*custom GUI based on MATLAB for systematic analysis of TFRAs. Currently, no interactive framework exists to detect the presence of HESs.*

The GUI integrates weight function visualization and automated single-sensor failure analysis within a unified platform. It enables verification of the double difference condition, detection of HESs, and interactive evaluation of candidate configurations, providing a simple, modular, and user-driven framework for validating functional robustness in sparse array design. Besides serving array designers in decision making, the proposed GUI could be instrumental in teaching complicated topics such as coarray redundancy, sensor essentialness, fragility and so on. Users can experiment with various array configurations and visualize how sensor positions influence array behavior.

### *1.4 Contributions*

Following are the main contributions of this work:

- A first-ever computational diagnostic tool capable of verifying the robustness of sparse arrays to single sensor failures has been developed. The underlying MATLAB codes used to design the core logic (backend) of the GUI are shared openly. Moreover, the simulator app is provided as an open research instrument via GitHub to facilitate reproducibility and future use.
- A comprehensive validation study has been performed on a few well-known sparse arrays to confirm the tool's functional accuracy. The proposed GUI tool acts as a litmus test that can tell apart nominal redundancy versus true functional robustness in any past, present, or future TFRAs.
- Thorough failure analysis of an existing class of TFRAs revealed that almost 50% of the array configurations suffer from the presence of HESs. A periodic pattern governing the occurrence of vulnerabilities has been found and reported for the first time.

The rest of the paper is organized as follows. Section 2 outlines the mathematical background and relevant sparse array terminology. Section 3 discusses the features of the proposed GUI and provides the core programming logic necessary to verify robustness. Section 4 deals with the procedure to design the frontend of the GUI. Section 5 presents numerical simulation results that confirm the functional accuracy of the tool and a specific source localization scenario whereby a TFRA can lose its DOA estimation ability under a single sensor failure. Section 6 presents a case study on the robustness analysis of existing TFRAs and reveals the hidden vulnerabilities and periodicities that violate the two-fold redundancy requirements. Section 7 concludes the paper with a suggestion for future enhancements.

## 2. Mathematical Background

This section reviews important sparse array terminology that is essential to understand the concepts described in the next few sections.

### 2.1 Sparse Array Terminology

#### 2.1.1. Difference Coarray, Holes, and Degrees of Freedom

Let $\mathbb{S}$ represent the physical array, with sensor positions normalized to half wavelength. The difference set $\mathbb{Z}$ lists all possible pairwise sensor separations in the physical array. Entries in $\mathbb{Z}$ denote spatial lags. The difference coarray (DCA) is formed by extracting non-repeating elements from $\mathbb{Z}$ and sorting them in the ascending order. The DCA is denoted by $\mathbb{D}$. Missing spatial lags create holes in the coarray. Sparse arrays with hole-free DCAs are preferred as they aid in unambiguous DOA estimation.

#### 2.1.2. Weight Function

The weight function $w(m)$ represents the number of sensor pairs with a separation distance $m$. For example, if an array has two sensor pairs, (2, 5), and (5, 8), we have $w(3) = 2$, as the separation distance in each pair of sensors is three.

*2.1.3. Desirable features of Two-Fold Redundant Arrays (TFRAs)*

Barring the failure of the first and the last sensor, an ideal TFRA shall withstand any other instance of single sensor failure without losing its hole-free DCA property. As per definition, for a TFRA with an aperture $L$, all the spatial lags except the last lag have a weight of two or more.

$$w(i) \geq 2; \; i \in [-L_u, L_u]$$

$$w(L) = 1 \tag{1}$$

where, $L_u = L - 1$.

*Essential Sensor:* An individual sensor is considered essential if its absence in the physical array alters the DCA span or continuity. By definition, the corner sensors of the array viz., 0 and $L$ are essential for any TFRSA.

*Hidden Essential Sensor:* Essential sensors occurring at positions other than 0 & $L$ can be categorized as hidden essential sensors (HESs).

*Fragility*: Number of essential sensors to the total number of sensors in an array. For a TFRSA, the expected fragility is $2/N$, as only the corner sensors are essential to preserve the DCA structure.

*2.2 Coarray MUSIC*

The signal model is based on coarray MUSIC which involves the Eigen decomposition of the coarray correlation matrix. The procedure has been exhaustively and extensively discussed in current literature [36], [37], [38], [39].

**3. Proposed Features of the GUI Tool**

While TFRSAs were theoretically designed to withstand single sensor failures, recent studies have identified the existence of hidden essential sensors (HESs) [29], [31] that prevent the arrays from achieving the desired functional robustness typically expected of TFRAs.

At present, there is no computational tool capable of performing systematic sensor failure analysis or verifying the functional resilience of a redundant array configuration. This gap necessitates the development of a dedicated software framework that can transition the focus from mere structural redundancy to verifiable robustness. The proposed GUI serves as a structured validation framework for analyzing TFRAs and consists of two core functional modules: (i) weight function computation and visualization, and (ii) single-sensor failure analysis. These modules are integrated within a unified interface to enable sequential evaluation of candidate array configurations, as shown in Fig. 1 below.

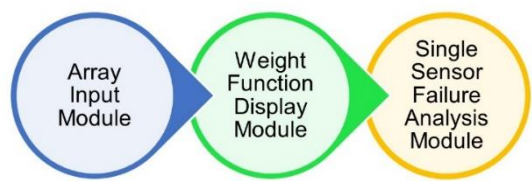

Fig. 1. GUI Functions

The input section allows the user to enter the test array in terms of sensor positions normalized to half wavelength. The weight distribution is displayed graphically to provide a direct representation of lag multiplicities across the aperture.

The MATLAB code for weight function computation and plotting is provided in the Appendix. Fig. 2 shows the options available for user navigation with respect to the array input and weight function module.

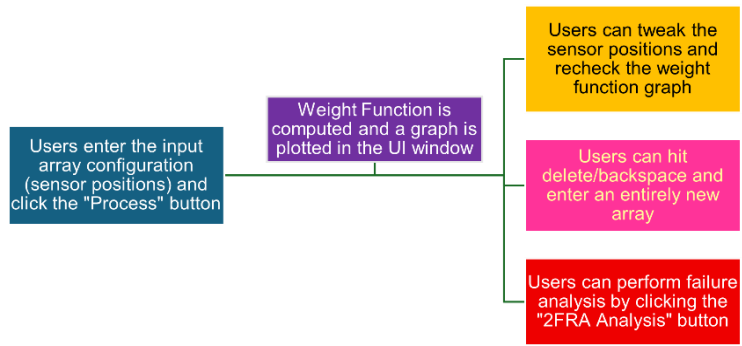

Fig. 2. User Interface for Input Section and Weight Function Plotting

The failure analysis module performs functional robustness validation through systematic single-sensor removal. It is noted that the first and last sensors (edge sensors) are inherently essential, as their removal necessarily alters the DCA. Therefore, the analysis focuses on interior sensors. Each interior sensor is removed sequentially, and the resulting DCA is recomputed. If the removal of any such sensor introduces holes, that sensor is classified as a hidden essential sensor (HES). Arrays containing interior HESs are deemed structurally fragile despite satisfying the double difference baseline condition (given in eq. (1)). Figure 3 outlines the procedure just described.

The output section consists of a dialog box that displays the characteristics of the test array from a two-fold redundancy lens as shown in Fig. 4. The tool classifies the input array into one of the three classes: (i) arrays without coarray redundancy which are not eligible for failure analysis, (ii) arrays with double difference baseline with hidden dependencies, and (iii) arrays with true two-fold redundancy.

1. Input the array to be tested. Initialize count←0.

2. Find the DCA and weight function in the healthy case.

3. Fail one sensor at a time from index 1 to index N-1 i.e., fail all sensors one-by-one except the first and the last sensor

4. Compute DCA and weight function of the resultant faulty array.

5. If the weight of any spatial lag is zero, increment the count by one and display the failed sensor considered in step 3.

6. Repeat steps 3-5 for all permitted single sensor failures.

7. If count is still zero, array has double redundancy. Otherwise, array contains hidden essential sensors (HESs) and is not a 2FRA.

Fig. 3. Core Programming Logic for Performing Robustness Analysis for backend of the GUI

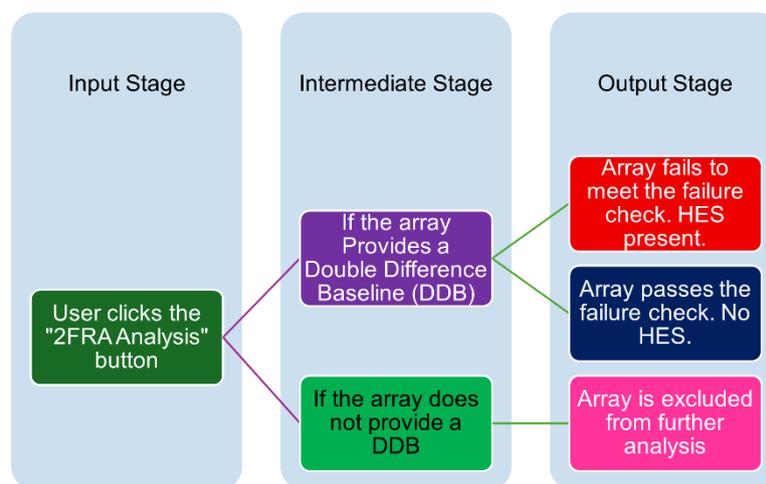

Fig. 4. Possible Scenarios During Robustness Analysis/Two-Fold Redundancy Check

By combining structural verification with explicit failure-based validation, we would like the GUI tool to filter out arrays that appear redundant but fail under sensor outage. We wish to design a reproducible and transparent framework for confirming true two-fold redundancy, making it a practical validation tool for robust sparse array design.

**4. GUI Design Methodology (Frontend)**

The GUI was developed using the MATLAB App Designer (matlab.apps.AppBase) following a structured and modular implementation. The layout was planned to reflect the workflow of two-fold redundancy analysis: array input, weight function plotting, and 2FRA validation.

*4.1 Array Input Section*

A text edit field (`uieditfield`) is used to accept the array configuration in a vector string format (e.g., [1 3 6 7]). The string input is converted into a numeric array using `str2num`. The configuration is internally normalized to a zero-based reference using `a = a - min(a)` before further processing. This ensures that the array is flushed either to the right or to the left to ensure that the first sensor always lies at the origin. As the difference coarray and weight function are immune to shifting, this does not cause any problems with the accuracy of results. In addition, it will provide flexibility to the tool in handling arrays with non-zero starting sensor positions. Validation checks ensure that the entries are non-negative integers, contain no duplicates (`numel(unique(a))`), and are properly formatted. Invalid inputs are handled through a dialog box using `uialert(app.UIFigure, ...)` to prevent execution of further steps.

*4.2 Weight Function Visualization Section*

A dedicated UIAxes component is used for graphical visualization. When the ProcessButton "`Plot Weight Function`" is clicked, the callback function computes the difference coarray (`a - a.'`), generates the indicator sequence `b`, and evaluates the weight function using `xcorr(b)`. The indicator sequence `b` is a binary string that represents the presence or absence of a sensor at the respective grid position. For example, the physical array with sensors at [0, 2, 5, 6] has the indicator sequence [1, 0, 1, 0, 0, 1, 1], indicating that the array does not have sensors at grid points $d, 3d,$ and $4d$. The resulting weight distribution is displayed as a stem plot using `stem (app.UIAxes, ...)`, with grid and axis labels configured within the same callback. The plotting area is visually separated from the control components to maintain clarity.

*4.3 Redundancy and Failure Analysis Section*

The AnalyzeButton "`Analyze for 2FRA`" triggers the 2FRA validation procedure through a separate callback function. The analysis is executed in three stages:

(i) verification of the double difference condition,

(ii) sequential single-sensor removal to evaluate fragility, and

(iii) identification of hidden essential sensors, if present.

The results are summarized and displayed in a dialog box using `uialert(app.UIFigure, ...)`. The plotting and analysis functionalities are implemented in separate callback functions (`ProcessButtonPushed` and `AnalyzeButtonPushed`) to maintain modularity and clarity of execution. The GUI is designed to be simple, interactive, and easy to use for testing different array configurations.

*4.4 Packaging and Installing the Application*

The core programming logic for weight function computation, plotting, and HES detection was validated through extensive preliminary checks before packaging the code as an executable application. The application (app) can be downloaded from the link (https://github.com/namya-malik/RMRA-TwoFold-Redundancy-Analyzer) and installed into the MATLAB app corner. Upon opening the app, the home screen shown in Fig. 5 appears in a new window. As seen, there is a text field to enter the array configuration of the AUT, two push buttons for call to action, one UI graphics element to plot the weight function. Not shown in Fig. 5 is a dialog box that opens upon clicking "Analyze for 2FRA".

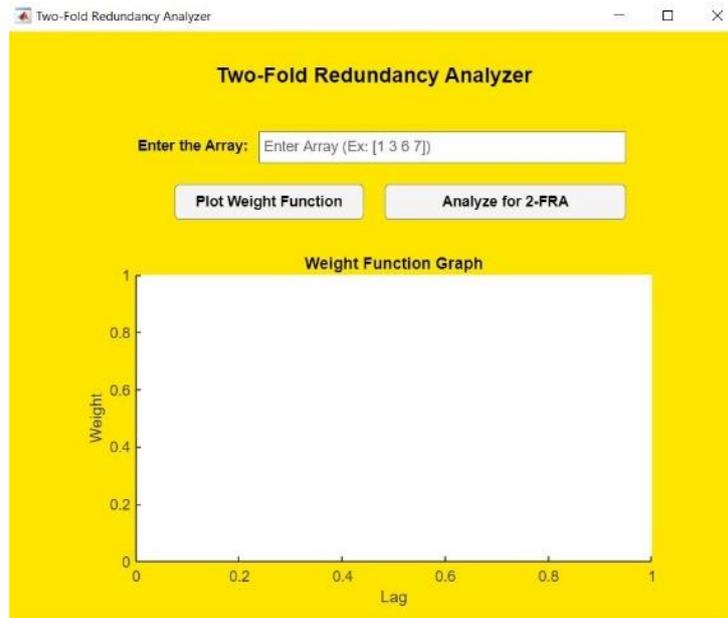

Fig. 5. Home Screen of the Designed GUI

## 5. GUI Validation/Numerical Results

We next show the output obtained from the GUI simulator under various scenarios. In particular, it is interesting to note how the tool interprets coarray redundancy and accurately classifies the input arrays into different categories based on the double difference criterion.

### 5.1 Weight Function Plotting

The tool can be used to visualize the weight function of any input array irrespective of whether it is uniform or sparse, nested or coprime, hole-free or not, and so on. Users can quickly assess the DCA properties such as missing spatial lags, hole-free property, double redundancy and so on, through visual means. Figure 6 shows the weight function plot for a 6-element MISC array with sensors at [0, 1, 2, 6, 10, 13]. It can be readily ascertained that the coarray is hole-free. The MISC array is chosen as a representative modern sparse array as it balances the achievable aperture and vulnerability to mutual coupling. The weight functions of many other arrays can be analyzed in a similar manner.

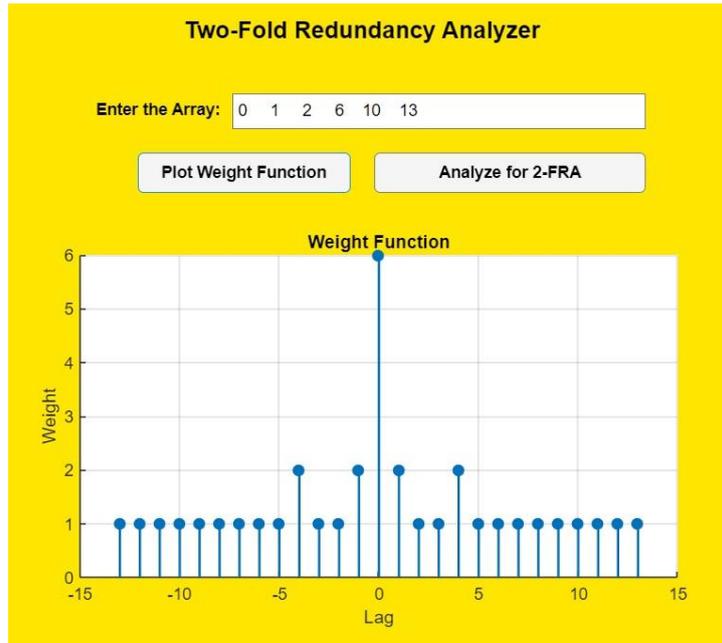

Fig. 6. Weight Function Graph of the 6-element MISC array

*5.2 Robustness Check (Single Sensor Failure Analysis and HES Detection)*

The main difference between the current GUI and one of the recently introduced tools for coarray analysis [35] is its ability to verify array robustness. While the recent tool can check DCA continuity and report the hole-free status of an array, it cannot perform failure mode analysis. We next verify the tool's functional accuracy by considering a few known sparse arrays from existing literature.

Arrays such as coprime, nested, MISC, and their variants can at most offer a hole-free DCA in the best case and do not possess double coarray redundancy. Hence, the GUI excludes all such arrays from further analysis. For example, consider the 6-element MISC array discussed in Fig. 6. When this array is given as the input to the "2FRA analyzer" module, we get the output shown in Fig. 7. The tool discards the array and aborts further analysis.

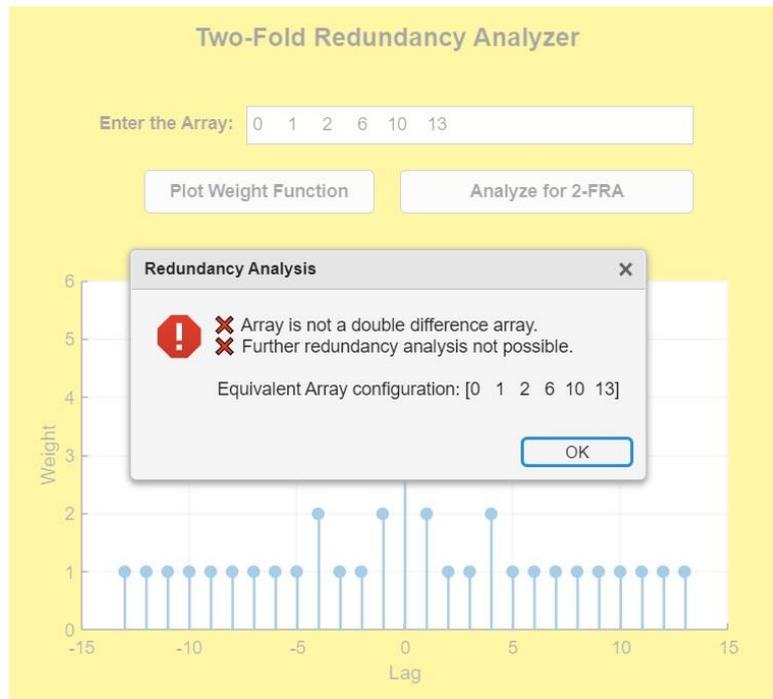

Fig. 7. GUI response when MISC array is given as an input to the robustness check module

*5.2.1 Double Difference Baselines*

Consider the 9-element 2FRA proposed by Zhu et al [28]. The array configuration in the interelement spacing (IES) notation is given by $\{1, 4, 1, 6, 1^4\}$. This corresponds to the sensor positions [0, 1, 5, 6, 12, 13, 14, 15, 16]. Fig. 8 shows the weight function of the above array. It is clear that the array has a doubly redundant DCA and satisfies eq. (1). However, upon subjecting this array to robustness check, the tool displays the presence of an HES at {6}, as shown in Fig. 9. To verify the correctness of the tool, we input the failed array (by removing the HES at {6}). The resulting weight function is as shown in Fig. 10. It can be seen that the failure of the sensor at position {6} indeed introduces a hole in the DCA at spatial lag 6. As this sensor does not lie at array edges, it can be classified as an HES. Hence, the above array does not qualify to be a TFRA in the true sense.

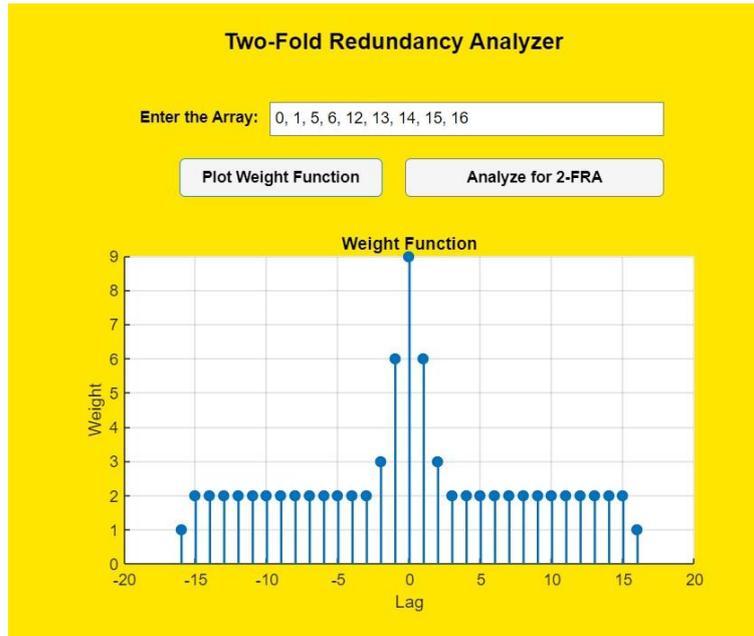

Fig. 8. Weight Function Graph of the 9-element 2FRA Proposed by Zhu et al

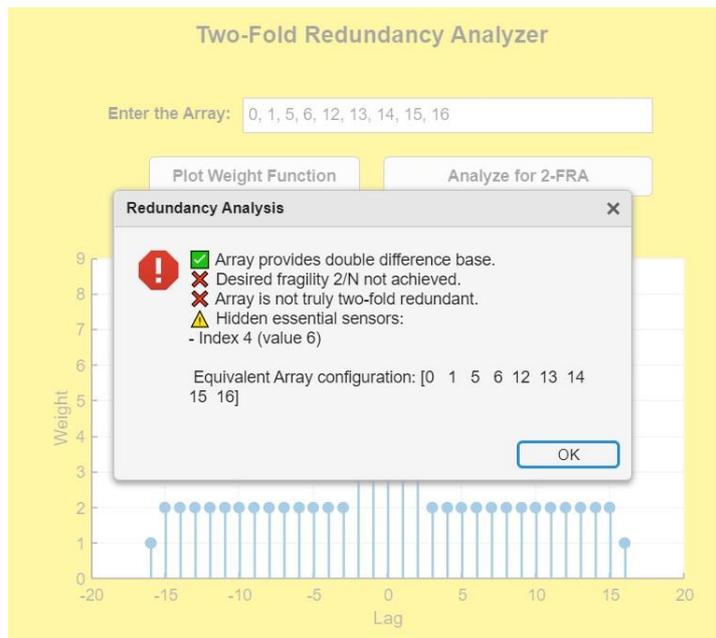

Fig. 9. Robustness Check of the 9-element 2FRA Against Single Sensor Failures

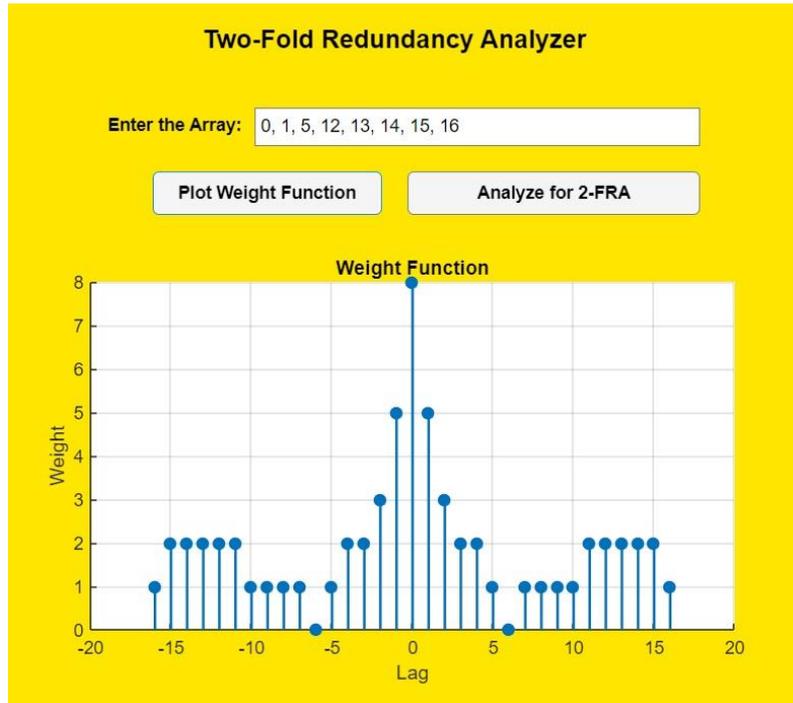

Fig. 10. Missing Spatial Lags in the DCA when the HES at {6} fails

It must be noted that operations such as shifting and flipping cannot sideline the hidden dependencies as the DCA is immune to operations such as translation (right or left shift) and/or reversal (flipping) of the physical array. For instance, the array [0, 1, 2, 3, 4, 10, 11, 15, 16] obtained after flipping the original 2FRA and subtracting 16 from each element (`sort(max(a)-a)`), suffers from a HES at {10}. As seen above, although some arrays appear to have doubly redundant DCAs, they might not be completely robust to single sensor failures due to the presence of hidden dependencies. Hence, it is important to ensure that TFRAs do not have any shared sensor redundancy.

*5.2.2 Truly Two-fold redundant*

Nevertheless, there are several TFRA classes that are truly robust. The symmetric nested array, the robust minimum redundancy array (RMRA), the ULA-inspired TFRA are all truly robust as they do not suffer from any hidden dependencies [26], [27], [29], [40]. For instance, consider the 9-element optimal RMRA found by Liu and Vaidyanathan. The robustness check yields the output that this array is truly robust and devoid of any HESs, as seen in Fig. 11.

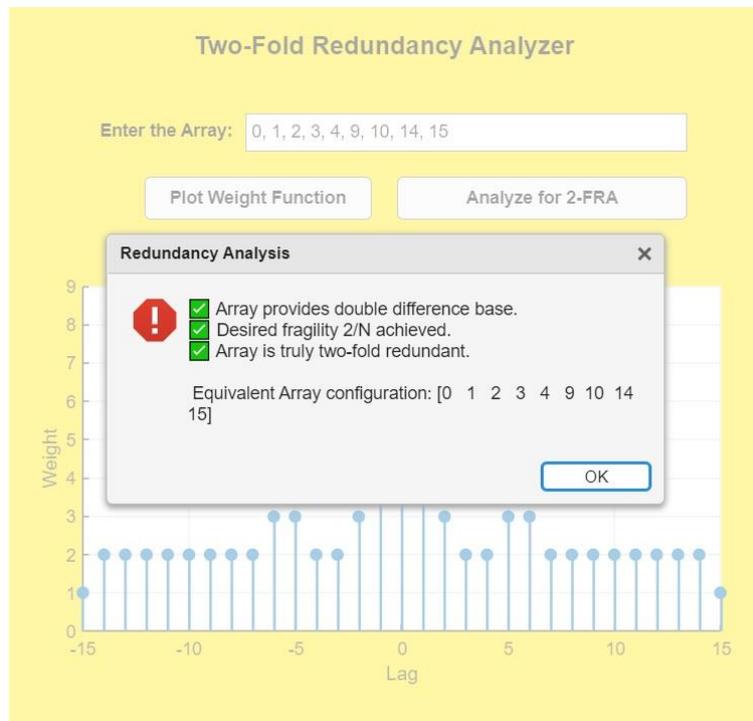

Fig. 11. Failure Analysis of a 9-Element TFRSA With True Two-Fold Redundancy

This validation confirms that the GUI can effectively differentiate between:

- Mere double difference bases, and
- Functionally robust two-fold redundant arrays.

### 5.3 Effect of HESs on DOA Estimation Accuracy

While the TFRA represents the general class of two-fold redundant arrays, we use the term 2FRA do denote a specific class proposed by Zhu et al. Consider the 13-element 2FRA with sensor positions [0, 1, 7, 8, 16, 17, 25, 26, 27, 28, 29, 30, 31]. It can be found from the GUI simulator that this array is a mere DDB and has a HES at {16}. We next show a numerical example on how the presence or rather the failure of a HES affects the array's DOA estimation capability.

Assume a DOA estimation scenario with eleven sources in the azimuth plane distributed evenly from -20° to 20° with an angular separation of 4°. Upon performing DOA estimation using coarray MUSIC on the signal received by the 13-element 2FRA during healthy and faulty cases, we obtained the pseudospectrum curves as shown in Fig. 12. It can be seen that the array performs accurate DOA estimation during the healthy case and during the failure of a non-HES (sensor at {17}). However, the failure of the HES at {16} breaks the coarray continuity and halves the available uniform degrees of freedom (DOFs). As a result, the array loses its ability to perform accurate DOA estimation and starts detecting false/ghost peaks while missing true targets. This erroneous behavior cannot be tolerated in real systems such as radar, sonar, radio astronomy etc. In Fig. 12, the term HES is denoted by SES which stands for "Surprise Essential Sensor".

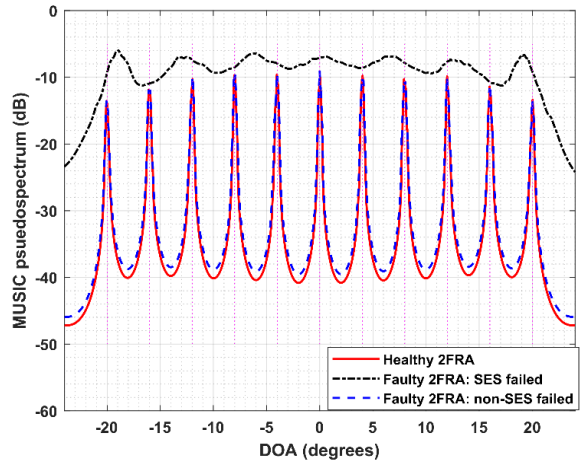

Fig. 12. DOA estimation results during healthy and faulty cases of the 13-element 2FRA

## 6. Case Study – Failure Analysis of Existing 2FRA Families

### 6.1 2FRA Formulation

The formulation of a $\beta$-FRA can be found in eq.(31)-eq.(33) of Zhu et al [3]. A 2FRA can be obtained by substituting $\beta = 2$. Given the number $N$ of sensors and 2FRA requirement, the optimal parameters $p, m$, for maximizing $L_u$ are given by

$$m^* = \left\lfloor \frac{N}{4} \right\rfloor$$
$$p^* = N - 2(m^* + 2) + 1 \quad (2)$$

where $\lfloor \cdot \rfloor$ denotes rounding to the nearest integer. The interelement spacing (IES) notation of the array is given by

$$\mathbb{S}_{MFRA^*} = \{1, p^*, (1, p^* + 2)^{m^*}, 1^{p^*}\} \quad (3)$$

### 6.2 2FRA Configurations and Failure Analysis

To test the robustness of existing 2FRAs, we created a MATLAB script to enumerate the array configurations using eq.(2) and eq.(3) above. 2FRA configurations shown in Table 1 from $N = 6$ to $N = 30$ were obtained by first computing the IES set using eq.(3) and then obtaining the sensor positions using the `cumsum` command available in MATLAB. Although the program can generate large arrays with sensor counts as high as 500, we have restricted the table only to 30, to conserve space. Each of these arrays was subjected to robustness test using the proposed GUI. The test revealed an interesting pattern in which almost half of the array configurations suffered from hidden dependencies, as shown in the last column of Table 1.

Table 1 List of 2FRA configurations from $N = 6$ to $N = 30$ and HES locations (if any)

| Array Size ($N$) | Array Configuration or Sensor Positions | HES position |
|---|---|---|
| 6 | [0 1 2 3 6 7] | {3} |
| 7 | [0 1 3 4 8 9 10] | {4} |
| 8 | [0 1 4 5 10 11 12 13] | {5} |
| 9 | [0 1 5 6 12 13 14 15 16] | {6} |
| 10 | [0 1 4 5 10 11 16 17 18 19] | {10} |
| 11 | [0 1 5 6 12 13 19 20 21 22 23] | {12} |
| 12 | [0 1 6 7 14 15 22 23 24 25 26 27] | {14} |
| 13 | [0 1 7 8 16 17 25 26 27 28 29 30 31] | {16} |
| 14 | [0 1 6 7 14 15 22 23 30 31 32 33 34 35] | Nil |
| 15 | [0 1 7 8 16 17 25 26 34 35 36 37 38 39 40] | Nil |
| 16 | [0 1 8 9 18 19 28 29 38 39 40 41 42 43 44 45] | Nil |
| 17 | [0 1 9 10 20 21 31 32 42 43 44 45 46 47 48 49 50] | Nil |
| 18 | [0 1 8 9 18 19 28 29 38 39 48 49 50 51 52 53 54 55] | {28} |
| 19 | [0 1 9 10 20 21 31 32 42 43 53 54 55 56 57 58 59 60 61] | {31} |
| 20 | [0 1 10 11 22 23 34 35 46 47 58 59 60 61 62 63 64 65 66 67] | {34} |
| 21 | [0 1 11 12 24 25 37 38 50 51 63 64 65 66 67 68 69 70 71 72 73] | {37} |
| 22 | [0 1 10 11 22 23 34 35 46 47 58 59 70 71 72 73 74 75 76 77 78 79] | Nil |
| 23 | [0 1 11 12 24 25 37 38 50 51 63 64 76 77 78 79 80 81 82 83 84 85 86] | Nil |
| 24 | [0 1 12 13 26 27 40 41 54 55 68 69 82 83 84 85 86 87 88 89 90 91 92 93] | Nil |
| 25 | [0 1 13 14 28 29 43 44 58 59 73 74 88 89 90 91 92 93 94 95 96 97 98 99 100] | Nil |
| 26 | [0 1 12 13 26 27 40 41 54 55 68 69 82 83 96 97 98 99 100 101 102 103 104 105 106 107] | {54} |
| 27 | [0 1 13 14 28 29 43 44 58 59 73 74 88 89 103 104 105 106 107 108 109 110 111 112 113 114 115] | {58} |
| 28 | [0 1 14 15 30 31 46 47 62 63 78 79 94 95 110 111 112 113 114 115 116 117 118 119 120 121 122 123] | {62} |
| 29 | [0 1 15 16 32 33 49 50 66 67 83 84 100 101 117 118 119 120 121 122 123 124 125 126 127 128 129 130 131] | {66} |
| 30 | [0 1 14 15 30 31 46 47 62 63 78 79 94 95 110 111 126 127 128 129 130 131 132 133 134 135 136 137 138 139] | Nil |

The periodic pattern shown above continued until $N = 500$, and probably beyond. Starting from $N = 10$, four arrays have HESs, four do not, and the same pattern repeats regularly. Hence, it can be concluded that almost 50% of 2FRA formulations suffer from the presence of an HES. As a result, these arrays cannot achieve true functional robustness required for practical DOA estimation or source localization under single sensor failures. If used blindly, these arrays could cause silent failures that can go unnoticed (missed targets or ghost peaks, as shown in subsection 5.3). Hence, it is important to formulate new classes of TFRAs safe from such hidden vulnerabilities and the proposed tool could go a long way in making that a reality. Apart from its use in research, the designed GUI can also be used as a teaching tool for explaining coarray redundancy, robustness to sensor failures, essential sensors, and fragility.

## 7. Conclusion & Future Scope

This paper presented the first-ever GUI tool capable of performing thorough failure analysis of two-fold redundant sparse arrays (TFRSAs). The concept of hidden essential sensors (HESs) was recently put forth, and this GUI provides a systematic platform for HES detection, thereby providing a practical instrument to visualize abstract concepts such as coarray redundancy, array robustness, essentialness of sensors in an array, and fragility. The simulator was designed from scratch, and its functional accuracy was validated by verifying its outputs against the properties of several well-known sparse arrays. Additionally, thorough failure analysis of a well-known class of 2FRAs revealed previously unknown hidden vulnerabilities and periodicities in failure patterns.

In the future, the functionality of the proposed GUI can be further enhanced to support additional structural diagnostics such as displaying the number of holes in the DCA so that users can immediately assess coarray continuity. The tool could also be extended to the 2D case to include thorough failure analysis of planar two-fold redundant arrays. It has to be noted that the tool is of great archival value as it can be used to verify the robustness of current as well as future TFRSAs.

**ACKNOWLEDGMENT**

We thank Vellore Institute of Technology, Vellore for providing campus-wide MATLAB license without which this study could not be completed.

**APPENDIX**

```
clc; clear all; close all;
%% Program to find DCA and weight function plot automatically. Just enter
the array configuration i.e., sensor positions
a = [0, 2, 3, 6, 14, 30, 35, 37, 39, 40] ; %Declare the physical array.
This works for ULAs and SLAs alike.
N = numel(a);
%% Following part of the code is to get the difference set, DCA, and weight
function.
x = a - a.'; ith column of x denotes a(i)-a. For example, 1st column is 0 -
{0, 2, 5, 8, 9}
d = reshape(x,[1 N*N]);
dca = unique(sort(d)); %This gives the DCA of the linear array.
w = histc(d,dca); %Alternate way to find weight function
stem(dca,w,'b','LineWidth',1.5);% xlabel('Spatial lags');
% ylabel('Weights');

if(length(dca)==(2*max(a)+1))
    disp('<strong> Coarray is hole-free </strong>')
else
    disp('<strong> Coarray has holes </strong>')
end
ylabel('Weights','FontSize',11)
xlabel('Spatial Lags', 'FontSize',11)
set(gca,'Xtick',[-40:10:40]);
ylim([0 numel(a)])
xlim([-max(a) max(a)])
grid on
grid minor
```